\documentstyle[12pt]{article}
\def\figcap{\section*{Figure Captions\markboth
        {FIGURECAPTIONS}{FIGURECAPTIONS}}\list
        {Figure \arabic{enumi}:\hfill}{\settowidth\labelwidth{Figure
999:}
        \leftmargin\labelwidth
        \advance\leftmargin\labelsep\usecounter{enumi}}}
 \relax
\newskip\humongous \humongous=0pt plus 1000pt minus 1000pt
\def\caja{\mathsurround=0pt}
\def\eqalign#1{\,\vcenter{\openup1\jot \caja
        \ialign{\strut \hfil$\displaystyle{##}$&$
        \displaystyle{{}##}$\hfil\crcr#1\crcr}}\,}
\newif\ifdtup

\def\jmp#1#2#3{J.\ Math.\ Phys.\ #1 (19#3) #2}

\def\np#1#2#3{Nucl.\ Phys.\ B#1 (19#3) #2}

\def\pr#1#2#3{Phys.\ Rev.\ D #1 (19#3) #2}
\def\prb#1#2#3{Phys.\ Rev.\ B #1 (19#3) #2}
\def\prep#1#2#3{Phys.\ Rep.\ #1 (19#3) #2}

\def\rmp#1#2#3{Rev.\ Mod.\ Phys.\ #1 (19#3) #2}

\newcounter{hran}
\def\bmini{\setcounter{hran}{\value{equation}}
\refstepcounter{hran} \setcounter{equation}{0}
\renewcommand{\theequation}{\thehran\alph{equation}}
              \begin{eqnarray}  }
\def\bminia{\setcounter{hran}{\value{equation}}
\refstepcounter{hran} \setcounter{equation}{1}
\renewcommand{\theequation}{\thehran\alph{equation}}
              \begin{eqnarray}  }
\def\bminiG#1{
          \setcounter{hran}{\value{equation}}
          \refstepcounter{hran}
          \setcounter{equation}{-1}
          \renewcommand{\theequation}{\thehran\alph{equation}}
          \refstepcounter{equation}
    \label{#1}
          \begin{eqnarray}          }
\def\emini{\end{eqnarray}\setcounter{equation}{\value{hran}}
\renewcommand{\theequation}{\arabic{equation}}}

\newskip\humongous \humongous=0pt plus 1000pt minus 1000pt
\def\caja{\mathsurround=0pt} \def\eqalign#1{\,\vcenter{\openup1\jot
\caja   \ialign{\strut \hfil$\displaystyle{##}$&$
\displaystyle{{}##}$\hfil\crcr#1\crcr}}\,} \newif\ifdtup

\def\half{\mbox{\small $\frac{1}{2}$}}

\def\ltap{\raisebox{-.4ex}{\rlap{$\sim$}} \raisebox{.4ex}{$<$}}

\def\frac#1#2{ {{#1} \over {#2} }}

\def\fun#1#2{\lower3.6pt\vbox{\baselineskip0pt\lineskip.9pt
  \ialign{$\mathsurround=0pt#1\hfil##\hfil$\crcr#2\crcr\sim\crcr}}}
\def\ie{\hbox{\it i.e.}{ }}      
      
\def\partder#1{{\partial   \over\partial #1}}

\def\re#1{(\ref{#1})}

\def\beq{\begin{equation}}
\def\eeq{\end{equation}}
\def\beeq{\begin{eqnarray}}
\def\eeeq{\end{eqnarray}}

\def\G{\Gamma}
\def\bG{ \bar \Gamma}

\def\L{ \Lambda}
\def\l{ \lambda}
\def\g{ \gamma}

\def\bp{ \bar p}
\def\d4#1{\frac {d^4 {#1} }{(2\pi)^4}}
\def\dLID#1{ \frac {\L \partial D^{-1}_{\L,\L_0}(#1)}{\partial \L}}

\def\dL{\L{\partial   \over\partial \L}}
\def\UV{$\L_0\to\infty\;$}
\def\IR{$\L\to 0\;\;$}

\def\bit{\begin{itemize}}
\def\eit{\end{itemize}}
\def\ben{\begin{enumerate}}
\def\een{\end{enumerate}}

\def\D{\Delta}
\def\Maxlp{\raisebox{-1.5 ex}{\rlap{\tiny $\;\;p_i^2<c\l^2$}}
\raisebox{0ex} {$\; \mbox{Max}\;\;\;\,$}}
\def\nome#1{{\label{#1}}}

\begin{document}
\begin{titlepage}
\renewcommand{\thefootnote}{\fnsymbol{footnote}}
\begin{flushright}
     UPRF 92-360  \\
     December 1992
\end{flushright}
\par \vskip 10mm
\begin{center}
{\Large \bf Perturbative renormalization and infrared \\
finiteness in the Wilson renormalization group: \\
the massless scalar case\footnote{Research supported in part by MURST,
Italy}}
\end{center}
\par \vskip 2mm
\begin{center}
        {\bf M.\ Bonini, M.\ D'Attanasio and G.\ Marchesini} \\
        Dipartimento di Fisica, Universit\`a di Parma and\\
        INFN, Gruppo Collegato di Parma, Italy
        \end{center}
\par \vskip 2mm
\begin{center} {\large \bf Abstract} \end{center}
\begin{quote}
A new proof of perturbative renormalizability and infrared finiteness for a
scalar massless theory is obtained from a formulation of renormalized field
theory based on the Wilson renormalization group. The loop expansion of the
renormalized Green functions is deduced from the Polchinski equation of
renormalization group. The resulting Feynman graphs are organized in such a
way that the loop momenta are ordered. It is then possible to analyse their
ultraviolet and infrared behaviours by iterative methods. The necessary
subtractions and the corresponding counterterms are automatically
generated in the process of fixing the physical conditions for
the ``relevant'' vertices at the normalization point.
The proof of perturbative renormalizability and infrared finiteness is
simply based on dimensional arguments and does not require the usual
analysis of topological properties of Feynman graphs.
\end{quote}
\end{titlepage}

\section{Introduction}
In perturbation theory renormalization requires a cumbersome
analysis of Feynman graph divergences and the exploitation of
topological properties in graph theory \cite{Bo}.
The origin of this complication is technical and due to the
overlapping divergence problem which arises in the procedure of
generating in a systematic way all appropriate subtractions.
This difficulty seems rather artificial since renormalization
is based on simple and general properties \cite{C} such as
dimensional analysis and the breaking of scale invariance in
interacting theories.
The perturbative study of finiteness of the vertex functions in a
massless theory \cite{LNK} brings a similar surprise.
Also in this case the reason for the finiteness is simple and general,
but in perturbation theory the analysis turns out to be rather
complex \cite{PQ}.

The deep physical meaning of renormalization can be appreciated
in the formulation due to Wilson \cite{W}-\cite{Be}.
One introduces a momentum cutoff $\L$, considers the effective
Lagrangian obtained by integrating the fields with frequencies
above $\L$ and
requires that the physical measures below such a scale are independent
of $\L$.
It is then natural to explore whether it is possible to deduce from
this formulation a proof of perturbative renormalizability which
avoids the usual cumbersome analysis of overlapping divergences.
This program has been successfully carried out by Polchinski \cite{P}.
{}From his formulation it is possible to deduce the perturbative expansion
and he was able to give a proof of perturbative renormalizability
for a scalar theory.
One does not need to control the detailed momentum dependence of
the effective couplings, thus the proof is based on simple dimensional
analysis.
However also this proof is technically quite laborious and the
physical features are not transparent. Moreover, although the proof is
perturbative, Green functions and Feynman graphs are never directly involved,
thus no connection with the usual analysis of perturbative renormalization
can be made.

In this paper we use the renormalization group equation of Ref.~\cite{P}
to obtain an alternative proof of perturbative renormalizability
for a scalar field theory.
Also in this case the proof is based on simple dimensional analysis but
in all steps we have an explicit connection to Feynman graphs.
Since power counting enters also in the infrared finiteness of the Green
functions at non exceptional momenta, we use similar
methods to obtain a perturbative proof of this feature in a massless theory.

We start by observing that the effective vertices satisfying the
renormalization group equations of Ref.~\cite{P}
correspond to the cutoff-Green functions of the theory in which
the parameter $\L$ plays the role of an infrared cutoff in the
propagators.
Thus the effective vertices evaluated at $\L=0$ correspond to
the renormalized Green functions.
The physical couplings, \ie the couplings with non negative
dimensions such as the mass, the wave function normalization, the
four point coupling, are fixed at the physical point $\L=0$.
In this way the renormalization group equations define in a constructive
way the renormalized quantum field theory.
This fact has been discussed also in Ref.~\cite{Be}-\cite{KK}.

We prove the renormalizability of the scalar theory by showing that by
fixing the physical couplings at $\L=0$ one generates,
order by order in perturbation theory, the necessary subtractions
which make finite any Green function.
The analysis of the ultraviolet behaviour of Feynman graphs does
not require
any complicated technical step so that the role of simple dimensional
counting and the origin of the needed subtractions are clear.
As in \cite{P}, we avoid the study of overlapping divergences.
Although we still refer to Feynman graphs we do not need a detailed
knowledge of their momentum dependence.
Solving by iteration the renormalization group equation, we find that
the Green functions are given as sum of contributions of Feynman
diagrams in which the various virtual momenta are ordered.
This allows us to control the ultraviolet behaviour by iterative methods.

In the same way we study the finiteness of the Green functions for
non exceptional momenta.
The perturbative analysis here is simpler since there is no need
to control the subtractions.
Due to the fact that the loop momenta are ordered, we need to
analyse only the infrared behaviour of the softest momentum.

The paper is organized as follows.
In Section~2 we rederive the renormalization group flow of Ref.~\cite{P}.
We show that the solution at $\L=0$ is the effective action
of the theory, deduce the perturbative expansion and perform
some one and two loop calculations. In Sections~3 and 4 we describe the proof
by induction of perturbative renormalizability and infrared finiteness.
Section~5 contains some remarks and conclusions. In the Appendix we
describe in this context the beta function.

\section{Renormalization group flow}
We consider a four dimensional massless scalar field theory
with a four point coupling $g$ at the scale $\mu$, \ie with
the two and four point vertex functions satisfying the conditions
\beq          \nome{norm}
\G_2(0)=0 \,,
\;\;\;\;\;\;\;\;
\frac {d \G_2(p^2)}{dp^2}\vert_{p^2=\mu^2}=1  \,,
\;\;\;\;\;\;\;\;
\G_4(\bp_1,\bp_2,\bp_3,\bp_4)=g  \, ,
\eeq
where $\bp_i$ are the momenta at the symmetric point
\beq\nome{sp}
\bp_i\bp_j=\mu^2(\delta_{ij}-\frac{1}{4}).
\eeq
The generating functional of Euclidean Green functions is
\beq  \nome{W}
Z[j]=e^{W[j]}=\int {\cal D}\phi\;
\exp
\left\{-\half\int \d4 p\phi(p)D^{-1}(p)\phi(-p)
\;{-S_{int}[\phi]}\;+ {\int \d4 p \phi(p)j(-p)}
\right\} \,,
\eeq
where $D(p)=1/p^2$ is the free propagator of the massless theory and
$ S^{int} [\phi]$ is the self interaction
\beq                 \nome{intaction}
S^{int}[\phi]=\half \int \d4 p
\left[ \g^{(B)}_2 p^2 +\g^{(B)}_3 \right]
\phi(p)\phi(-p) +
\frac {\g^{(B)}_4}{4!} \int d^4x \Phi^4(x) \, ,
\eeq
with $\phi(p)$ the momentum representation of the field $\Phi(x)$.
Feynman graphs  generated by \re{W} and \re{intaction} have ultraviolet
divergences, which can be regularized by
assuming that the fields have frequencies bounded by
an ultraviolet cutoff $\L_0$ ($p^2<\L_0^2$) or, equivalently, that the
propagator $D(p)$ vanishes for $p^2 > \L_0^2$.
The parameters $\g^{(B)}_2$ and $\g^{(B)}_4$ are dimensionless,
$\g^{(B)}_3$ has the dimensions of a square mass, and they are fixed by the
physical conditions \re{norm}.

\subsection{Derivation of the evolution equation}
The evolution equation which is the basis for the present treatment is
derived as follows.
We consider the functionals $Z[j;\L]$ and $W[j;\L]$ obtained
from \re{W} in which we replace $D(p)$ by the cutoff propagator
\beq \nome{prop}
D(p) \Rightarrow D_{\L,\L_0}(p)= \frac { K_{\L,\L_0}(p)}{p^2} \,,
\eeq
with $K_{\L,\L_0}(p)=1$ in the region
$\L^2 \, \ltap \,p^2\, \ltap \, \L_0^2$
and rapidly vanishing outside.
As one takes the limit $\L \to 0$, the cutoff Green functions
generated from $W[j;\L]$ tend, at least perturbatively, to the
physical ones, \ie we have $W[j;\L] =W[j]$ for \IR.

The functional $Z[j;\L]$ can be computed from the following
evolution equation
\beq\nome{evZ}
\dL Z[j;\L]= - \half \int \d4 q \dLID {q} \, (2\pi)^8
            \frac { \delta^2 Z[j;\L]}
                  {\delta j(q)\delta j(-q)}\,,
\eeq
obtained by differentiating \re{W} with the propagator $D_{\L,\L_0}(p)$.
As we will discuss in detail later, the boundary conditions
are determined by requiring that the vertex functions satisfy
the normalization conditions \re{norm} at $\L=0$, and the form
\re {intaction} at $\L=\L_0$. Since we are interested in the limit
\UV, in these functionals we do not write explicitly the dependence
on the ultraviolet cutoff.
Notice that the inverse cutoff propagator entering in \re{evZ} is cancelled by
corresponding propagators in the external legs of the Green functions.
Although inverse cutoff propagators cancel, in this
section we shall assume a cutoff function different from zero everywhere
but rapidly vanishing outside the range $(\L,\L_0)$.

The present treatment is related to the Wilson renormalization
formalism. Actually, Eq.~\re{evZ} is equivalent to
the renormalization group equation of Ref. \cite{P}.
To clarify this relation recall that in the Wilson formulation
one integrates \re{W} over the fields with frequency larger than $\L$
to obtain a new action $S_{\L}[\phi_{\L},j]$ and Eq.~\re{W} becomes
$$
Z[j]=e^{W[j]}=\int {\cal D} \phi_{\L} \;
\exp
\left\{ -S_{\L}[\phi_{\L},j] \; \right\} \, ,
$$
where $\phi_{\L}(p)$ are the soft fields with frequency smaller
than $\L$.
Obviously $Z[j]$ does not depends on $\L$, and this is the basis
of Wilson renormalization group.
In particular we can take $\L$ small.
If we consider sources $j(p)$ with frequencies larger than $\L$, the
integration over $\phi_{\L}$ generates only virtual loops with
soft momenta $p^2 < \L^2$.
Since the theory is infrared finite (at least perturbatively,
see Section 4), these contributions vanish for $\L \to 0$
and in this limit we can ignore the $\phi_{\L}$ dependence in the
new action and we can write
$$
S_{\L} [\phi_{\L},j]  \simeq S_{\L} [0,j]= -W[j;\L] \to -W[j] \, .
$$
The two functionals $S_{\L} [0,j]$ and $-W[j;\L]$ are equal,
since to use the cutoff propagator $D_{\L,\L_0}(p)$
is equivalent to integrate only over fields with frequencies larger
than $\L$.
In the framework of renormalization group, it is more typical to
consider sources with frequencies softer than $\L$ and compute
$S_{\L}[\phi_{\L},0]$ which is the action at the scale $\L$.
It is easy to show that $S_{\L}[\phi_{\L},0]=S_{\L}[0,j]$ with
$\phi_{\L}(p)=D_{\L,\L_0}(p)j(p)$. This explains why the
Polchinski equation for $S_{\L}[\phi_{\L},0]$ is the same as
the evolution
equation for $W[j;\L]$ obtained from \re{evZ} provided one identifies
$\phi_{\L}(p)$ with $D_{\L,\L_0}(p)j(p)$.
The fact that $S_{\L}[\phi_{\L},0]$ generates cutoff Green functions with
amputated external propagators has been observed also
in Refs.~\cite{Be}-\cite{KK}.

\subsection{Evolution equation for the vertex functions}

To study the renormalizability and the infrared finiteness of this
theory it is more convenient to consider the proper vertices
$\Gamma_{2n}(p_1,\cdots,p_{2n})$ (see Fig.~1a) and their generating functional
$$
\G[\psi]=
\sum_{n=1} \frac 1 {(2n)!} \int \prod_{i=1}^{2n} \d4 {p_i} \psi(p_i)
\, \G_{2n}(p_1, \cdots,p_{2n})
\, (2\pi)^4\delta^4(\sum_{i=1}^{2n} p_i) \; ,
$$
which is related to $W[j]$ by the Legendre transformation
$$
\G[\psi]=-W[j] + W[0]+\int \d4 p j(p)\psi(-p) \, ,
\;\;\;\;\,\;\;
 (2\pi)^4  \frac {\delta \G}{\delta \psi(-p)} =j(p) \, .
$$
In the same way we introduce the cutoff effective action
$\G[\psi;\L]$ as Legendre transformation of $W[j;\L]$. Taking into account
that
$$
\dL \G[\psi;\L]=-\dL W[j;\L]+\dL W[0;\L] \,,
$$
the evolution equation for the cutoff effective action is given by
\beq \nome{ev1}
\eqalign{
\dL \biggr\{ &\G[\psi;\L]-\half \int \d4p D^{-1}_{\L,\L_0}(p)\,\psi(p)\psi(-p)
\biggr\}
\cr&
=\half \int \d4q \dLID {q} \, (2\pi)^8
\frac { \delta^2 W[j;\L]}{\delta j(q)\delta j(-q)}
+\dL W[0;\L]\;.
}
\eeq
The functional in the integrand is the inverse of
$\delta^2\G[\psi]/\delta\psi(q)\delta\psi(q')$
\beq\nome{inv}
\delta^4(q+q')=(2\pi)^{8}
\int d^4 {q''}
\frac{\delta^2 \Gamma[\psi]}{\delta\psi(q)\delta\psi(q'')}
\frac{\delta^2 W[j]}{\delta j(-q'')\delta j(q')}\;,
\eeq
and is obtained as follows.
First we isolate the two point function contribution in the
two functionals
\beq\nome{inv2}
\eqalign{
(2\pi)^8\frac{\delta^2\G[\psi]}{\delta\psi(q)\delta\psi(q')}
=&
(2\pi)^4\delta^4(q+q') \G_2(q;\L)+
\G_2^{int}[q,q',\psi]\,,
\cr(2\pi)^8\frac {\delta^2\, W[j;\L]}{\delta j(q)\delta j(q')}
=&
(2\pi)^4\delta^4(q+q')\frac{1}{\G_2(q;\L)}+
W_2^{int}[q,q',j]\,.
}
\eeq
Notice that the two point function contribution in the second
equation cancels the constant term $\L\partial W[0;\L]/\partial\L$ in
\re{ev1}.
Then, from \re{inv} we obtain $W_2^{int}$ as functional of $\psi(p)$
$$
W_2^{int}[q,q',j]=-
\frac {1}{\G_2(q;\L)}
\bG[q,q',\psi]
\frac {1}{\G_2(q';\L)} \,,
$$
where the auxiliary functional $\bG$ satisfies the equation
\beq \nome{bGrec}
\bG[q,q',\psi] = \G_2^{int}[q,q',\psi]-
\int \d4 {q''} \G_2^{int}[q,-q'',\psi]
\frac{1}{\G_2(q'';\L)}
\bG[q'',q',\psi]   \, .
\eeq
By expanding this equation we obtain the auxiliary vertices
$\bG_{2n+2}(q,p_1, \cdots,p_{2n},q';\L)$ (see Fig.~1b)
in terms of the proper vertices. For $n=1$ we find
$$
\bG_4(q,p_1,p_2,q';\L)=
 \G_4(q,p_1,p_2,q';\L)\,,
$$
and in general (see Fig.~2)
\beq\nome{auxv}
\eqalign{
&
\bG_{2n+2} (q,p_1,\cdots,p_{2n},q';\L)=
\G_{2n+2} (q,p_1,\cdots,p_{2n},q';\L)
\cr&
-\sum
\G_{2k+2} (q,p_{i_1},\cdots,p_{i_{2k}},-Q;\L)
\;\frac{1}{\G_2(Q;\L)}
\bG_{2n-2k+2}(Q,p_{i_{2k+1}},
\cdots,p_{i_{2n}},q';\L)
}
\eeq
where $Q=q+p_{i_1}+\cdots p_{i_{2k}}  $ and
the sum is over $k=1 \dots n-1$ and over the ${2n\choose 2k}$
combinations of $(i_1 \cdots i_{2n})$.

In conclusion the evolution equation for the functional $\G[\psi]$ is
$$
\eqalign{
\dL \biggr\{ \G[\psi;\L]
&-\half \int \d4p D^{-1}_{\L,\L_0} (p)\,\psi(p)\psi(-p) \biggr\}
\cr
&= - \half \int \d4q
\dLID {q}\biggr[ \frac {1}{\G_2(q;\L)} \biggr]^2\,
\bG[q,-q,\psi]\, .
}
$$
After isolating the interaction part of the two point function
$$
\G_2(p;\L)=D^{-1}_{\L,\L_0}(p)+\Sigma(p;\L)\,,
$$
the evolution equation for the proper vertices are
\bminiG{eveq}
\nome{eveq1}
\dL
\Sigma(q;\L)
=\half \int \d4 q
\frac{S(q;\L)}{q^2}
\G_{4}(q,p,-p,-q;\L)\,,
\emini
and
\addtocounter{equation}{-1}
\bmini
\refstepcounter{equation}
\nome{eveq2}
\dL
\G_{2n}(p_1,\cdots,p_{2n};\L)
=\half \int \d4 q
\frac{S(q;\L)}{q^2}
\bG_{2n+2}(q,p_1,\cdots,p_{2n},-q;\L)\,,
\emini
where $S(q;\L)$ is given by
\beq \nome{esse}
\frac{S(q;\L)}{q^2}
\equiv
{ \frac {\L \partial D_{\L,\L_0}(q)}{\partial \L}}
\left[ \frac {1}{1+D_{\L,\L_0}(q)\Sigma(q;\L)} \right]^2\,.
\eeq
These equations involve in the right
hand side vertices at the infrared cutoff $\L$ with a pair of
exceptional momenta $q$ and $-q$.
For \IR these vertices become singular since we are dealing with a
massless theory.
In a next Section we analyse this limit and show that Eq.~\re{eveq}
allows one to obtain vertex functions $\G_{2n}$ with non exceptional
momenta which are finite for \IR, order by order in perturbation theory.

\subsection{Physical couplings and boundary conditions}
In the study of renormalization we should control the
quantities with non negative momentum dimension, \ie the
``relevant'' couplings. Dimensional analysis gives
$$
\G_{2n} \sim \bG_{2n} \sim \L^{4-2n}\,.
$$
Thus the ``relevant'' couplings are
$$
\g_2 (\L) = \frac {d\Sigma(p^2;\L)}{dp^2}\vert_{p^2=\mu^2} \,,
\;\;\;\;\;\;\;\;
\g_3 (\L) =\Sigma(0;\L)\,,\;\;\;\;\;\;\;\;
\g_4 (\L) =\G_4(\bp_1,\bp_2,\bp_3,\bp_4;\L)\,,
$$
and correspond, for $\L=0$, to the physical couplings introduced in
(\ref{norm}).
We then isolate the relevant couplings in the two and
four point vertices
\beq\nome{isol}
\eqalign{
\Sigma(p^2;\L)&= p^2 \g_2(\L) + \g_3 (\L) +
\Delta_2(p^2;\L) \,, \cr
\G_4(p_1,p_2,p_3,p_4;\L)&= \g_4(\L)+\Delta_4(p_1,p_2,p_3,p_4;\L)\,,
}
\eeq
where $\Delta_2(0;\L)=0$, $\partial \Delta_2(p^2;\L)/\partial p^2=0$
at $p^2=\mu^2$, and $\Delta_4(\{\bp_i\};\L) =0 $
at the symmetric point \re{sp}.
{}From dimensional analysis we have
$$
\eqalign{
&
\g_2 \sim \g_4 \sim (\L)^0\;, \;\;\;\;\;\; \g_3 \sim (\L)^{2}\,,
\cr&
\Delta_2\sim \Delta_4 \sim (\L)^{-2}\,.
}
$$
Notice that in $\Delta_2(p^2;\L)$ four powers of momentum are
absorbed by the $p$-dependence required by the two conditions at $p^2=0$
and $p^2=\mu^2$. Similarly in $\D_4$ two powers of momentum are absorbed
by the $p_i$-dependence required by the condition at the symmetric
point.

To obtain the vertex functions from the evolution equation
(\ref{eveq}) we need the boundary conditions.
For the relevant couplings $\g_i(\L)$ they are defined at the physical
value $\L=0$ by the conditions \re{norm}
\bminiG{bound}
\nome{bound1}
\g_2(\L=0) =0\;,\;\;\;\;
\g_3(\L=0) =0\;,\;\;\;\;
\g_4(\L=0) =g\;.
\emini
The physical requirement we have to set on the remaining vertex
functions is that they are irrelevant when the ultraviolet cutoff
$\L_0$ is set to infinity.
The simplest choice is to set all these ``irrelevant'' vertices to zero
at $\L=\L_0$
\addtocounter{equation}{-1}
\bmini
\refstepcounter{equation}
\nome{bound2}
\Delta_2(p^2;\L_0) =0 \,,
\;\;\;\;\;
\Delta_4(p_1,p_2,p_3,p_4;\L_0)=0\,,
\;\;\;\;\;
\G_{2n}(p_1,\cdots,p_{2n};\L_0)=0 \,,
\;\;\;\;\; n \ge 3\,.
\emini
With these conditions the functional $\G[\psi;\L]$ has the form of the
action in (\ref{intaction}) with $\g_i^{(B)}$ given by the
relevant couplings $\g_i$ evaluated at $\L=\L_0$.
The bare coupling constant is then
$ g^{(B)}= {\g_4^{(B)} }/{(1+\g_2^{(B)})^2 }$.

The evolution equations (\ref{eveq}) with the boundary conditions
\re{bound}, can be converted into a set of integral
equations.
For the three relevant couplings $\g_i$ the boundary conditions
(\ref{bound1}) give
\beq\nome{inte1}
\eqalign{
\g_{2}(\L)&=
\half \int \d4 q \int_0^{\L}
\frac{d\l}{\l} \frac{S(q;\l)}{q^2}
\frac{\partial}{\partial p^2}
\G_{4}(q,p,-p,-q;\l)|_{p^2=\mu^2}\,,
\cr
\g_{3}(\L)&=
\half \int \d4 q \int_0^{\L}
\frac{d\l}{\l} \frac{S(q;\l)}{q^2}
\G_{4}(q,0,0,-q;\l)\,,
\cr
\g_{4}(\L)=g &+
\half \int \d4 q \int_0^{\L}
\frac{d\l}{\l} \frac{S(q;\l)}{q^2}
\bG_{6}(q,\bp_1,\ldots,\bp_{4},-q;\l)\,.
}
\eeq
For the other vertices, the boundary conditions (\ref{bound2}) give
\beq\nome{inte2}
\eqalign{
\G_{2n}(p_1\ldots p_{2n};\L)&=
-\half \int \d4 q \int_{\L}^{\L_0}
\frac{d\l}{\l} \frac{S(q;\l)}{q^2}
\bG_{2n+2}(q,p_1,\ldots,p_{2n},-q;\l)\,,
\cr
\Delta_{2}(p;\L)&=
-\half \int \d4 q \int_{\L}^{\L_0}
\frac{d\l}{\l} \frac{S(q;\l)}{q^2}
\Delta \G_{4}(q,p,-p,-q;\l)\,,
\cr
\Delta_{4}(p_1 \ldots p_4;\L)&=
-\half \int \d4 q \int_{\L}^{\L_0}
\frac{d\l}{\l} \frac{S(q;\l)}{q^2}
\Delta \bG_{6}(q,p_1,\ldots,p_{4},-q;\l)\,,
}
\eeq
where $n>2$ in the first equation.
The subtracted vertices $\Delta \G_{4}$ and $\Delta \bG_{6}$ are defined by
\beq \nome{inte3}
\eqalign{
\Delta \G_{4}(q,p,-p,-q;\l)
\equiv
&
\G_{4}(q,p,-p,-q;\l) -\G_{4}(q,0,0,-q;\l)
\cr
&
-p^2 \frac{\partial}{\partial p'^2}
\G_{4}(q,p',-p',-q;\l)|_{p'^2=\mu^2}\,,
\cr
\Delta \bG_{6}(q,p_1,\ldots,p_{4},-q;\l)
\equiv
&
\bG_{6}(q,p_1,\ldots,p_{4},-q;\l)
-\bG_{6}(q,\bp_1,\ldots, \bp_{4},-q;\l)\,.
}
\eeq
The subtractions in $\Delta \G_{4}$ and $\Delta \bG_{6}$ are a consequence
of isolating in eq.~\re{isol} the relevant couplings in the two and four
vertices and of the different boundary conditions \re{bound}.
As we expect they provide the necessary subtractions to
make finite the vertex functions for $\L_0\to\infty$ at any order
in perturbation theory.

We should notice the different role of boundary conditions for the
relevant couplings at $\L=0$ and of the irrelevant vertices at
$\L=\L_0 \to \infty$. For the relevant couplings this implies that
the $q$-integration is bounded above by $\L$.
This is crucial for obtaining a finite result since, as expected from
dimensional counting, the integrands grow with $q^2$.
The bare couplings, obtained by setting $\L=\L_0$, are therefore
growing with $\L_0$ and give the counterterms of the Lagrangian
\re{intaction} in terms of the physical coupling $g$.
For the other vertices the $q$-integration is bounded above by the
ultraviolet cutoff $\L_0$. To show that the theory is renormalizable, one
must prove that for $q^2 \to \infty$ the vertices in the integrands in
\re{inte2} are sufficiently vanishing to allow one to take \UV.

For a sharp cutoff Eq.~\re{esse} can be written
\beq \nome{esse1}
\frac{S(q;\l)}{q^2} = - \frac {1}{\l} \delta ( \l-\sqrt{q^2} ) s(\l)\,,
\;\;\;\;\;\;
s(\l)=\left[ \frac{1}{1+\frac{1}{\l^2}\Sigma(\l;\l)} \right]^2 \,,
\eeq
which is independent of $\L_0$.

\subsection {Loop expansion}

By solving iteratively Eqs.~\re{inte1} and \re{inte2}
one obtains the loop expansion.
Here we compute the first terms for $\L_0 \to \infty$ so that
the cutoff propagator is simply
$$
D_{\L}(q)=K_{\L}(q)/q^2\,,
$$
where $K_{\L}(q)=1$ for $q^2\geq \L^2$ and vanishes for $q^2<\L^2$.
It will be clear that the limit $\L\to 0$ can be taken only for non
exceptional momenta.
The starting point is the zero loop order in which  the only non
vanishing vertex is
$$
\g_4^{(0)} (\L) =g\,,
$$
and the auxiliary vertices with $n \ge 2$ are given by (see Fig.~3)
\beq \nome{bG2n+20l}
\bG_{2n+2}^{(0)}(q,p_1,\cdots,p_{2n},q';\L)=
-(-g)^n \; \sum_{perm} \prod_{k=1}^{n-1}D_{\L}
 \left( q+\sum_{\ell=1}^{2k}p_{i_\ell} \right) \,,
\eeq
where the sum is over $(2n)!/2^n$ terms obtained from the
permutations of $(p_{i_1},\cdots, p_{i_{2n}})$ and the symmetry of
the four point coupling.

\vskip .3 true cm \noindent
1. {\it One loop vertices.}
\vskip .2 true cm \noindent
The only non vanishing contribution for the two point function is
$$
\g_3^{(1)}(\L) = \half g \int\d4 q \D_{\L}(q)
= - \frac{1}{32\pi^2}  g \L^2\,,
$$
where
$$
\D_\L(q)=D_{\L}(q) - D_{0}(q)
= -\frac{1}{q^2}\theta(\L^2-q^2) \,.
$$
{}From \re{bG2n+20l} one obtains
$$
\g^{(1)}_4(\L)\,  =\, -\frac 3 2 g^2\int\d4 q
\left\{ D_{\L}(q) D_{\L}(q+\bp)-
       D_{0}(q) D_{0}(q+\bp)
\right\}\,,
$$
where $\bp=\bp_i+\bp_j,i \neq j$, and we used the symmetry of the
subtraction point \re{sp}.
For large $\L$ the range of integration is bounded by $q^2 \ltap \L^2$
and we get
$$
\g_4^{(1)}(\L) \simeq \frac{3}{16\pi^2} g^2\ln(\L/\mu)\,, \;\;\;\;\;
\mu \ll \L \,.
$$
For small $\L$ we have
$$
\g_4^{(1)}(\L) \sim \L^2\,\;\;\;\;\L \ll \mu\,.
$$
The remaining part $\Delta_4$ of the four point vertex is obtained in a
similar way
$$
\eqalign{
\Delta_4^{(1)}(p_1,\cdots,p_4;\L) =& -\half g^2\int\d4 q D_{\L}(q)
\cr &
\times
\left\{ D_{\L}(q+p_1+p_2)+\cdots -3D_{\L}(q+\bp) \right\}\,,
}
$$
where the dots stand for the other two terms with $p_2$ replaced by
$p_3$ and $p_4$.
Due to the subtractions the integration is convergent for $q^2 \to \infty$.
For large infrared cutoff this integral vanishes as $\mu^2/\L^2$
and $P^2/\L^2$ with $P$ a combination of external momenta.
The physical value is obtained at $\L=0$ and one has
\beq \nome{onel1}
\G_4^{(1)}(p_1,\cdots,p_4) = \frac{1}{32\pi^2} g^2
\left\{ \ln[\frac{(p_1+p_2)^2}{\mu^2}]+\cdots \right\}
\,.
\eeq
In the Appendix we use this results to obtain, in this formulation,
the one loop beta function.

For the vertices $\G^{(1)}_{2n}$ with $n\ge 2$ we have
\beq \nome{onel2}
\G_{2n}^{(1)}(p_1,\cdots,p_{2n};\L)=
- \frac {(-g)^n} {2n} \int\d4 q \, D_\L(q) \,
\; \sum_{perm} \prod_{k=1}^{n-1}D_{\L}
 \left( q+\sum_{\ell=1}^{2k}p_{i_\ell} \right)\,.
\eeq
The integral is convergent for large $q^2$.
At the physical value $\L=0$ these vertex functions become
singular for vanishing momenta.
However it is known \cite{CW} that the effective potential obtained by
summing the vertex functions for vanishing momenta is infrared finite.
We rederive here the one loop effective potential $V(\psi)$ in order to
illustrate in this framework the role of the regularization and the
physical conditions in \re{norm}.
Apart from a volume factor, $V(\psi)$ is given by $\G[\psi]$ obtained
with the source $\psi(p)=(2\pi)^4\delta^4(p) \psi$.
For non vanishing $\L$ we get
$$
V(\psi)=\half \g_3^{(1)}\,\psi^2
+\frac {1}{4!}(g+\g_4^{(1)})\, \psi^4 -\half \int \d4  q \Theta(q^2-\L^2)
\left\{ \sum_{n=2}^{\infty} \frac 1 n \left(\frac {-g\psi^2}{2q^2} \right)^n
- \frac {(g\psi^2)^2}{8 q^2(q+\bar p)^2} \right\}\,.
$$
The various terms diverge at $q=0$ for
$\L=0$. However, performing the sum and then taking $\L=0$, we have
$$
V(\psi)= \frac {g}{4!}\psi^4 + \half \int \d4  q
\left\{ \ln  \left(1+\frac{g\psi^2}{2q^2}\right)
-\frac{g\psi^2}{2q^2}
+\frac{(g\psi^2)^2}{8 q^2(q+\bar p)^2}
\right\}\,.
$$
This expression does not have any infrared singularity for $q=0$
and the integral is convergent at large $q$ (see Ref.~\cite{CW}).


\vskip .3 true cm\noindent
2. {\it Two loop propagator.}
\vskip .2 true cm\noindent
By using the previous results we find
\beq \nome{two2}
\eqalign{
& \G_2^{(2)}(p;\L) =
\G_2^{(2)}(p;0)-
\frac{g^2}{3!} \int\d4 q \D_\L(q) \int\d4 {q'}
\biggr\{ \D_\L(q')\D_\L(q+q'+p)
\cr&
+
\frac{3}{2}\biggr[ D^2_{\L}(q')
+2D_{0}(q')D_{\L}(q+q'+p)
-3D_{0}(q')D_{0}(q+q'-\bp)\biggr]\biggr\} \,.
}
\eeq
The $q$-integration is bounded by the factor $\D_\L(q)$ which gives
$q^2\leq\L^2$.
The $q'$-integration for the first term in the curly bracket
is also bounded by $\L$.
For the second term the $q'$-integration is convergent because of
subtractions.
The value of $\G_2^{(2)}(p;0)$ is given by
$$
\G_2^{(2)}(p,0)=-\frac{g^2}{3!}\int\d4 q\int\d4 {q'}
\frac{1}{q^2 q'^2}\biggr[\frac{1}{(q+q'+p)^2}-\frac{1}{(q+q')^2}
-p^2   \frac{\partial}{\partial p'^2} \frac{1}{(q+q'+p')^2}\biggr]
$$
where $p'^2=\mu^2$ and the integrations are convergent due to the
subtractions.

\section{Perturbative renormalizability}
In this section we prove that the theory is perturbative renormalizable,
namely that in \re{inte2}  we can set \UV.
As shown before, the loop expansion is obtained by iterating
Eqs.~ \re{inte1} and \re{inte2}.
{}From the vertices $\G_{2n}^{(\ell)}$ one constructs the integrands at
the loop $\ell$ which give the next loop vertices upon $q$-integration.
The convergence of the integrals giving $\G^{(\ell+1)}_{2n}$
will be ensured by dimensional counting, while the one for
$\Delta_2$ and $\Delta_4$ will require the subtractions in
$\Delta \G_4$ and $\Delta \bG_6$ given in \re{inte3}.
The convenient way to represent the subtracted vertices $\Delta \G_4$
and $\Delta \bG_6$ is by Taylor expansion as for
the Bogoliubov $R$ operators.
Since we are interested in the large $\l$ behaviour we use the
expansion around vanishing momenta.
We need to consider only even derivatives since odd derivative
terms vanish for symmetry either directly or after integration.
The subtracted vertex $\D\bG_6$ is obtained considering the expansion
\bminiG{taylor}
\nome{t1}
& \bG_6(q,p_1,\cdots, p_4,-q;\l) = \bG_6(q, 0,\cdots, 0,-q;\l) \nonumber \\
& +\int_0^1dx(1-x)\left(\sum_{i=1}^3 p_i \cdot \partial'_{i,4}\right)^2
\bG_6(q,p'_1,\cdots, p'_4,-q;\l)|_{p'_i=xp_i}
\emini
where $\partial'_{i,4}= {\partial}/{\partial p'_i}-
{\partial}/{\partial p'_4}$.
The first term, which is the most singular contribution, is
cancelled in the subtracted quantity $\Delta \bG_6$.
For $\D\G_4$ we need to consider the expansion up to four derivatives
\addtocounter{equation}{-1}
\bmini
\refstepcounter{equation}
\nome{t3}
& \D\G_4(q,p,-p,-q;\l)=
\frac 1{3!}\int_0^1dx(1-x)^3(p \cdot \partial')^4
\G_4(q,p',-p',-q;\l)|_{p'=xp} \nonumber \\
& -\half (p \cdot \partial')^2
\left\{ \G_4(q,p',-p',-q;\l)|_{p'^2=\mu^2}-
       \G_4(q,p',-p',-q;\l)|_{p'=0} \right\}
\emini
where $\partial'=\partial/\partial p'$.
Notice that also the second term can be expressed by the fourth derivative
of $\G_4$. Similarly the integrand for $\g_2$ can be expressed in terms of
the second derivative of $\G_4$ with respect to the momentum components.


In order to prove that the theory is perturbatively renormalizable we
have to analyse the behaviour for large $\l$ of the vertices in the
integrands and show that the integration over $\l$ is convergent for
$\l\to\infty$ (the convergence of the integrals for $\l\to 0$ will be
discussed in the next section).
In this analysis we are not interested in the detailed dependence
of the vertices on the external momenta, except
for the fact that the integration momentum is fixed at $q^2=\l^2$
(see \re{esse1}).
To prove perturbative renormalizability it will be sufficient,
as in \cite{P}, to bound the large $\l$ behaviour of the vertices in
which all external momenta do not exceed the cutoff.
We then introduce the following functions which depend only on $\l$
\beq\nome{norma}
|f_{2n}|_{\l}\equiv
\Maxlp
|f_{2n}(p_1,\cdots, p_{2n};\l)|
\eeq
where $c$ is some numerical constant and $f(p_1,\cdots, p_n;\l)$ is
$\G_{2n}$, $\bG_{2n+2}$ or one of their derivatives.
Iterating \re{inte1} and \re{inte2} in which we set \UV,
we obtain the following bounds.
For the relevant couplings
\bminiG{e1}
\nome{e1a}
\g^{(\ell+1)}_2(\L)
&\ltap& \int_0^{\L^2}d\l^2s^{(\ell-\ell')}(\l)
|\partial^2 \G^{(\ell')}_4|_{\l} \;, \\
\nome{e1b}
\g^{(\ell+1)}_3(\L)
&\ltap& \int_0^{\L^2}d\l^2s^{(\ell-\ell')}(\l)
|\G^{(\ell')}_4|_{\l} \;, \\
\nome{e1c}
\g^{(\ell+1)}_4(\L)
&\ltap& \int_0^{\L^2}d\l^2s^{(\ell-\ell')}(\l)
|\bG^{(\ell')}_6|_{\l} \;.
\emini
For the irrelevant vertices
\bminiG{e2}
\nome{e2a}
|\G^{(\ell+1)}_{2n}|_{\L}
&\ltap& \int_{\L^2}^{\infty }d\l^2s^{(\ell-\ell')}(\l)
|\bG^{(\ell')}_{2n+2}|_{\l} \;, \\
\nome{e2b}
|\Delta_2^{(\ell+1)}|_{\L}
&\ltap& \L^4 \int_{\L^2}^{\infty }d\l^2s^{(\ell-\ell')}(\l)
|\partial^4 \G^{(\ell')}_{4}|_{\l} \;, \\
\nome{e2c}
|\Delta_4^{(\ell+1)}|_{\L}
&\ltap& \L^2 \int_{\L^2}^{\infty }d\l^2s^{(\ell-\ell')}(\l)
|\partial^2 \bG^{(\ell')}_{6}|_{\l} \; .
\emini
For the derivatives of vertices
\bminiG{e3}
\nome{e3a}
|\partial^m\G_{2n}^{(\ell+1)}|_{\L}
&\ltap& \int_{\L^2}^{\infty} d\l^2 s^{(\ell-\ell')}(\l)
|\partial^m\bG_{2n+2}^{(\ell')}|_{\l} \; , \\
\nome{e3b}
|\partial^m\Delta_2^{(\ell+1)}|_{\L}
&\ltap& \L^4\,\int
_{\L^2}^{\infty} d\l^2 s^{(\ell-\ell')}(\l)
|\partial^{m+4}\Delta_4^{(\ell')}|_{\l} \; , \\
\nome{e3c}
|\partial^m\Delta_4^{(\ell+1)}|_{\L}
&\ltap& \L^2\,\int_{\L^2}^{\infty} d\l^2
s^{(\ell-\ell')}(\l)|\partial^{m+2}\bG_6^{(\ell')}|_{\l} \; .
\emini
where $\partial^m$ stands for $m$ partial derivatives with respect to
momentum components and the factors $\L^2$ and $\L^4$ in front of
integrals come by maximizing the $p^2$ or $p^4$ factors in
\re{t1} and \re{t3} respectively. Actually $\partial^m$ in \re{e3b}-\re{e3c}
could also act on these $p$ factors. As we will show in the following all
these contributions are of the same order.

We now prove by induction that the theory is perturbatively renormalizable,
namely that the integrals in \re{e2} and \re{e3} are convergent for
$\l \to \infty$.
By using the majorization in \re{e1}-\re{e3}, as one can expect, the
proof becomes very simple.

\noindent
1) {\it Assumptions at the loop $\ell$.}

\noindent
The assumptions we make are simply given by dimensional counting except for
logarithmic corrections and they involve the nine quantities above.

\noindent
a) Relevant couplings ($T=\log(\L/\mu)$)
\beq \nome
 {a1}
\g^{(\ell)}_2(\L)= {\cal O}(T^{\ell-1}),\;\;\;\;
\g^{(\ell)}_3(\L)= {\cal O}(\L^2 T^{\ell-1}),\;\;\;\;
\g^{(\ell)}_4(\L)= {\cal O}(T^{\ell})\,.
\eeq
\noindent
b) Irrelevant vertices
\beq \nome {a2}
|\G_{2n}^{(\ell)}|_{\L}={\cal O}(\L^{4-2n}T^{\ell-1}),\;\;\;\;
|\Delta_2^{(\ell)}|_{\L}={\cal O}(\L^{2}T^{\ell-2}),\;\;\;\;
|\Delta_4^{(\ell)}|_{\L}={\cal O}(T^{\ell-1})\,,
\eeq
\noindent
c) Derivative vertices
\beq \nome {a3}
|\partial^m\G_{2n}^{(\ell)}|_{\L}={\cal O}(\L^{4-2n-m}T^{\ell-1}),\;\;\;\;
|\partial^m\Delta_2^{(\ell)}|_{\L}={\cal O}(\L^{2-m}T^{\ell-2}),\;\;\;\;
|\partial^m\Delta_4^{(\ell)}|_{\L}={\cal O}(\L^{-m} T^{\ell-1})\,.
\eeq
These assumptions are satisfied for $\ell=0$ and $1$.

\noindent
2) {\it Iteration to the loop $\ell+1$.}

\noindent
Notice that the powers of $\L$ in \re{a1}-\re{a3} are independent of the
number of loops since they are dictated by dimensional counting.
For the relevant (irrelevant) couplings the integrands increase (decrease)
with $\l$ thus the integrals are dominated by the upper (lower) limit
$\l=\L$. For the irrelevant couplings we can therefore take the limit
\UV, removing the ultraviolet cutoff.
It is simple to see that the integrals in \re{e1}-\re{e3}
reproduce at loop $\ell+1$ the same dimensional counting behaviours.
This is just what
is needed to prove the perturbative renormalizability, since logarithmic
corrections cannot change the power counting at any finite order.
Actually it is relatively simple to control also the powers of $T$
and in the following we show that the behaviours \re{a1}-\re{a3} are
reproduced by the iteration.

Before discussing the large $\L$ behaviours at the $\ell+1$ loop we
derive from \re{a1}-\re{a3} some intermediate results for the
integrands at loop $\ell$.

\noindent
a) From the two point function and \re{esse1} we have
$$
s^{(\ell)}(\l)\sim t^{\ell-1} \,,
$$
where $t\equiv\log(\l/\mu)$.

\noindent
b) The leading term of the auxiliary vertices $\bG_{2n+2}$
is given by the contribution of Fig.~4 in which only four point vertices
are involved
$$
|\bG_{2n+2}^{(\ell)}|_{\l}\sim\l^{2-2n}\prod_1^n\g_4^{(\ell_i)}(\l)
\sim\l^{2-2n}t^{\ell}\,,
$$
where $\sum\ell_i=\ell$ and we have a factor $\l^{-2}$ for each internal
propagator. All other contributions coming from higher vertices and from
loop corrections in the intermediate propagators give the same
power in $\l^2$ but a lower power in $t$.

\noindent
c) The leading term of the derivatives of the auxiliary vertices
is again obtained from the contribution of Fig.~4
in which the derivatives act on the internal propagators
$$
|\partial^m\bG^{(\ell)}_{2n+2}|_{\l}\sim\l^{2-2n-m}t^{\ell}\,.
$$
Again, the contributions from derivatives of higher vertices or from
loop corrections of internal propagators give lower powers of $t$.

By using these results in \re{e1}-\re{e3} we reproduce at $\ell+1$ loop
order the behaviours in \re{a1}-\re{a3}
In all cases we have $\ell'=\ell$; \ie loop corrections of the
propagator in $s(\l)$ are not contributing to the leading terms.

\section{Infrared behaviour}
In this section we show that for the massless scalar theory,
the vertex functions at non exceptional momenta are finite order by
order in perturbation theory.
Namely we prove that the integration over $q$ in \re{inte1} and \re{inte2}
is convergent at the lower limit when we take $\L \to 0$.
As in the case of renormalizability, this is shown by induction on the
number of loops.

The integrands in \re{inte1} and \re{inte2} are given by vertices
with one pair $(q, -q)$ of exceptional momenta. Thus, by iteration,
one introduces vertices with any number of pairs of
exceptional momenta.
In general we say that a pair of momenta $p_i$ and $p_j$
in $\G_{2n}(\{p_i\};\L)$ is exceptional if $p_i+p_j={\cal O}(\L)$.
In the following we add an index to the vertex functions
to identify the number of pairs of exceptional momenta. We write
$$
\G_{2n,s}(\{p_i\};\L) \equiv \G_{2n}(\{p_i\};\L) \,,
\;\;\;\;\mbox {for}\;\;\;\;
(p_{i_1}+p_{i_2}) , \cdots ,
(p_{i_{2s-1}}+p_{i_{2s}})= {\cal O}(\L)\,.
$$
where $s=0,\ldots,n-1$.
For $s=n-1$ all pairs of momenta are exceptional.
Therefore we denote by $\G_{2n,0}$ the vertices  without exceptional
momenta. Similar notation will be used for the auxiliary vertices
$\bG_{2n+2,s+1}$ with $s+1$ pairs of exceptional momenta.

\vskip 0.3 cm \noindent
1) {\it Assumptions at the loop $\ell$}
\vskip 0.2 cm \noindent
For $\L \to 0$ we assume the following behaviours
\bminiG{asx}
\nome{as1}
\G_{2n,0}^{(\ell)}(\L) &\to& \mbox{finite} \,, \\
\nome{as2}
\G_{2n,1}^{(\ell)}(p_1, \cdots, p_{2n} ;\L) &=& {\cal O}( T^{\ell}) \,, \\
\nome{as3}
\G_{2n,s}^{(\ell)}(p_1,\cdots, p_{2n}; \L) &=& {\cal O}( \L^{2-2s}T^{\ell-1})
\, \;\;\;\; s=2 \cdots n-1\,, \\
\nome{as4}
\G_{2}^{(\ell)}(p;\L) &=& {\cal O}( \L^2\,T^{\ell-1})
\, \;\;\;\; \mbox{for} \; p^2= {\cal O}( \L^2)\, ,  \\
\nome{as5}
\frac {\partial}{\partial p_{\mu}}
\G_{2n,n-1}^{(\ell)}(p,-p,p_1,\cdots, p_{2n-2}; \L)
&=& {\cal O}( \L^{4-2n}T^{\ell-1} \; \frac {p_{\mu}+P_{\mu}}{\L^2} )
\, ,
\emini
where $T=\ln(\L/\mu)$.
The first equation states the most important result.
In \re{as5} all momenta are of order $\L$ and $P_{\mu}$
stands for any combination of $p_i$'s.
These assumptions are satisfied for $\ell=0$ and $1$ (see Subsection
2.3).

\vskip 0.3 cm \noindent
2) {\it Iteration at loop $\ell+1$}
\vskip 0.2 cm \noindent
Before discussing the $\L \to 0$ behaviour at loop $\ell+1$ we derive from
\re{asx} some intermediate results for the vertices at loop
$\ell$ entering in the integrands of \re{inte1} and \re{inte2}

\noindent
a) From the two point function we have
\beq\nome{ci1}
s^{(\ell)}(\l) = {\cal O}(t^{\ell-1})\,,
\eeq
where $t=\log(\l/\mu)$.

\noindent
b) For the four point function we have
$$
\frac{\partial }{\partial p^2}
\G_{4,1}^{(\ell)}(q,p,-p,-q;\l)|_{p^2=\mu^2,\,q^2=\l^2}
={\cal O} ( t^{\ell})\,,
$$
$$
\Delta \G_4^{(\ell)}(q,p,-p,-q;\l)|_{q^2=\l^2} = {\cal O}(t^{\ell}) \,.
$$
The first equation is obtained from \re{as2} observing that the derivative is
evaluated at non vanishing momentum, while the second is
obtained from the behaviour of the single terms in
the subtracted vertex $\Delta\G_4$. Actually there are cancellations
giving a less singular behaviour, but we do not need to analyse them.

\noindent
c) For the auxiliary vertices with $s=0$ we have
\beq\nome{ci6}
\bG^{(\ell)}_{2n+2,1}(q,p_1,\cdots,p_{2n},-q;\l)|_{q^2=\l^2}\sim
\G_{2n+2,1}^{(\ell)}(\l)\sim t^{\ell} \,.
\eeq
This behaviour is given by the first contribution of Fig.~2.
Since only the pair $(q,-q)$ is exceptional, the other terms
involve vertices with non exceptional momenta which are finite.

\noindent
d) For the auxiliary vertices with $s>0$ we have
\beq\nome{ci10}
\bG_{2n+2,s+1}^{(\ell)}(q,p_1,\cdots,p_{2n},-q;\l)|_{q^2=\l^2}\sim
\l^{-2s}t^{\ell}\,,\;\;\;\;\;\;\;s\ge 1\,.
\eeq
This behaviour is obtained by taking the largest number of
internal propagators at momentum $q^2=\l^2$.
This is given by the contribution of Fig.~5 in which the $s$
pairs of exceptional momenta
among the $p_i$'s are emitted in the four point functions to
the left (or right). In this way we have $s$ internal propagators at
momentum $q^2=\l^2$ giving the factor $\l^{-2s}$ in \re{ci10}.
Loop corrections to the internal propagators give non leading
logarithmic powers.

\noindent
e) For the derivatives of the auxiliary vertices with all
exceptional momenta we have
\beq\nome{ci13}
\frac {\partial}{\partial p_{\mu}}
\bG_{2n+2,n}^{(\ell)}(q,p,-p,p_1\cdots p_{2n-2},-q; \l)
={\cal O}( \l^{4-2n}t^{\ell}\; \frac {p_{\mu}+q_{\mu}+ P_{\mu}}{\l^4} )
\,,\;\;\;\;\;\;n>1\,.
\eeq
This behaviour is given by the graphs of Fig.~6. Here $P_{\mu}$ is
a combination of $p_i$'s and the derivative acts on
a propagator connecting the two vertices with $p$ or $-p$ incoming.
All other contributions with higher vertices, with loop corrections
in the propagators or with the derivatives acting on vertices lead to
lower powers of $t$.

We now deduce the $\L \to 0$ behaviour of the $\ell+1$ vertices.

\vskip 0.3 true cm \noindent
a) For the two point function we find
$$
\g_2^{(\ell+1)}(\L) = {\cal O}(\L^2 T^{\ell}) \to 0 \, ,
$$
$$
\g_3^{(\ell+1)}(\L)
= {\cal O}(\L^2 T^{\ell}) \to 0 \,,
$$
$$
\Delta_2^{(\ell+1)}(p;\L) = {\cal O}(\L^{0}) \, .
$$
One can show that the behaviour for $\g_2^{(\ell+1)}$ can be
improved to $\L^2 T^{\ell-2}$.
These results show that $\G_2^{(\ell+1)}$ is finite for $\L \to 0$,
thus verifying the main assumption in \re{as1} for the case $n=1$.

To study the behaviour of $\D_2^{(\ell+1)}$ for $p^2={\cal O}(\L^2)$
we use the Taylor expansion
\beq\nome{ci51}
\eqalign{
\D_2^{(\ell+1)}(p;\L)= &
\half\int_{\L} \d4 q \frac{s^{(\ell-\ell')}(\l)}{q^2}
\cr &
\times
\int_0^1 dx (p \cdot \partial')
\G_{4,1}^{(\ell')}(q,p',-p',-q;\l)|_{\l^2=q^2,\; p'=xp}
={\cal O}(p^2T^{\ell}).
}
\eeq
This proves \re{as4} at loop $\ell+1$.
All these behaviours  are obtained by ignoring loop corrections in
$s(\l)$ since they would give lower powers of $T$.

\vskip 0.3 true cm \noindent
b) For the four point function at non exceptional momenta
we obtain from \re{ci6}
\beq\nome{ci7}
\g_4^{(\ell+1)}(\L) = {\cal O}(\L^2 T^{\ell}) \to 0 \,,\;\;\;\;\;\;
\Delta_4^{(\ell+1)}(p_1,\cdots,p_4;\L)
= {\cal O}(\L^{0}) \,,
\eeq
where again for the leading terms we do not have contributions from
loop corrections in $s(\l)$.

For $\G_{2n,0}^{(\ell+1)}$ with $n>2$ we obtain from \re{ci6}
\beq\nome{ci9}
\G_{2n,0}^{(\ell+1)}(p_1,\cdots,p_{2n};\L)={\cal O}(\L^{0})\,.
\eeq
We conclude from \re{ci7}-\re{ci9} that all physical vertices
at $\L=0$ and non exceptional momenta
are finite.
Moreover, as required by the condition $\g_4 (0)=g$,
the relevant coupling $\g_4^{(\ell+1)}(\L)$ vanishes
at $\L=0$.
This verifies at loop $\ell+1$ the main assumption \re{as1}.

\vskip 0.3 true cm \noindent
c) For the vertices with pairs of exceptional momenta
we prove \re{as2}-\re{as3} at the $\ell+1$ loop order
by using the result in \re{ci10}.


\vskip 0.3 true cm \noindent
d) For the vertices with all exceptional momenta
we prove \re{as5} at the $\ell+1$ loop order.
For $n=1$ this is simply obtained by taking the derivative of
Eq.~\re{ci51} with respect to $p_{\mu}$.
For $n>1$ this can be done by writing
$$
\frac {\partial}{\partial p_{\mu}}
\G_{2n,n-1}^{(\ell+1)}(p,-p,p_1,\cdots, p_{2n-2}; \L)
$$
$$
=\half\int_{\L} \d4 q \frac{s^{(\ell-\ell')}(\l)}{q^2}
\frac {\partial}{\partial p_{\mu}}
\bG_{2n+2,n}^{(\ell')}(q,p,-p,p_1,\cdots, p_{2n-2},-q; \l)|_{\l^2=q^2}
$$
and using \re{ci13}.

\section{Final comments}

We have used the renormalization group technique to obtain in a constructive
way equations for
the renormalized vertices of a scalar massless theory. The loop expansion
is derived from the iterative solution of these equations
and is an expansion in the physical coupling $g$.
In this approach subtractions come from the fact that we isolated
(see Eq.~\re{isol}) in the two and four point functions the three relevant
vertices $\g_i(\L)$ for which we have to impose the boundary conditions
\re{norm} at the physical value $\L=0$.
This procedure allows us to generate in a systematic way the subtracted
vertices $\D\G_4$ and $\D\bG_6$ in \re{inte3} which, after integration,
give $\D_2$ and $\D_4$, the ``irrelevant'' part of the two and four
point function.
These subtracted vertices and the corresponding counterterms obtained
from $\g_i(\L)$ evaluated at $\L=\L_0$, can be expanded in the
physical coupling $g$.
This procedure of constructing the subtracted vertices is very similar to
the one of Bogoliubov $R$-operators \cite{Bo} but it seems to be based
on a more physical ground.
Once the subtracted vertices are identified, the proof of
renormalizability depends only on dimensional counting.

In this formulation the simplifications of the analysis of the
ultraviolet and infrared behaviour of the integrands of Feynman graphs
is based on the fact that, at a given loop $\ell+1$, the momentum $q$
in \re{inte1} and \re{inte2} sets the cutoff for the vertices at the
previous loop. The difference between relevant and irrelevant couplings
is in the range of $q$-integrations with respect to the cutoff $\L$.
We have then that the vertices are given by combinations of Feynman graphs
in which the internal loop momenta are ordered.
This ordering depends on whether we are generating a relevant or
irrelevant vertex. It is clear that once we have an ordering in the
momenta, the asymptotic behaviour of Feynman graphs contributions can be
studied by iterative procedure.

\vspace{3mm}\noindent{\large\bf Acknowledgements}

We have benefited greatly from discussions with C.\ Becchi and
G.\ Korchemsky.

\eject
\newpage

\begin{center}
{\Large \bf Appendix
}\end {center}

We discuss in the present formulation the role of the subtraction
point $\mu$ and deduce the beta function (see also Ref.~\cite{H}).
Denote by $\G_{2n}(g,\mu)= \G_{2n}(p_1,\cdots,p_{2n};g,\mu)$
the vertices defined by \re{inte1} and \re{inte2} which satisfy the
physical conditions \re{norm} with coupling $g$ at the scale $\mu$.
We assume the theory renormalizable and infrared finite so we set
$\L=0$ and \UV. Therefore $\mu$ is the only dimensional parameter,
apart for the external momenta.
However physical measurements should not depend on the specific
value of $\mu$.
Consider the vertices $\G_{2n}(g',\mu')$ with coupling $g'$ at a
new scale $\mu'$.
The request that the two sets of vertices  $\G_{2n}(g,\mu)$
and $\G_{2n}(g',\mu')$ describe the same theory
implies that the corresponding effective actions
$\G[\psi;g,\mu]$  and $\G[\psi';g',\mu']$ are equal, where the two
fields $\psi$ and $\psi'$ are related by a rescaling,
$\psi'(p)=\sqrt{Z_2}\psi(p)$.
This implies that the two sets of vertices are related by
$$
\G_{2n}(g',\mu')=Z_2^{-n} \G_{2n}(g,\mu) \,.
$$
Notice that the evolution equations \re{eveq} are invariant with respect to
this rescaling, provided we rescale also the propagator
$D_{\L,\L_0}(p)$ accordingly.
This implies, as usual, that $Z_2$ is fixed only by the physical
condition \re{norm}.
To obtain $Z_2$ we use the decomposition \re{isol}
$$
\G_2(p;g,\mu)=p^2+\Delta_2(p;g,\mu)\,,
$$
where
$$
\Delta_2(0;g,\mu)=0\,, \;\;\;\;\;
\frac{\partial}{\partial p^2}\Delta_2(p;g,\mu)|_{p^2=\mu^2}=0\, ,
$$
as required by the physical conditions \re{norm}.
We use the corresponding decomposition and normalization for the
vertex $\G_2(p;g',\mu')$.
The renormalization function $Z_2$ is obtained
by expanding $Z_2\G_2(p;g',\mu')=\G_2(p;g,\mu)$ at $p^2=\mu^2$ or
$p^2=\mu'^2$. We find
$$
Z_2=1+a_2(\mu';g,\mu) \,=\, \frac {1}{1+a_2(\mu;g',\mu')} \,,
$$
where $a_2$ is the dimensionless quantity
$$
a_2(\mu';g,\mu) \equiv
\frac{\partial}{\partial p^2}\Delta_2(p;g,\mu)|_{p^2=\mu'^2}\, .
$$
Similarly we use the decomposition \re{isol} for the four point function
$$
\G_4(p_1,\cdots,p_4;g,\mu)=g+\Delta_4(p_1,\cdots,p_4;g,\mu)\,,
$$
where $\Delta_4(\bp_1,\cdots,\bp_4;g,\mu)=0$  at the symmetric point
in \re{sp}.
Again we use the corresponding decomposition and normalization for the
vertex $\G_4(p_1,\cdots,p_4;g',\mu')$.
Introducing the dimensionless quantity
$$
a_4(\mu';g,\mu) \equiv \Delta_4(\bp'_1,\cdots,\bp'_4;g,\mu)\,,
\;\;\;\;\;
\bp'_i\bp'_j=\mu'^2(\delta_{ij}-\frac 14)\;,
$$
from $Z^2_2\G_4(g',\mu')=\G_4(g,\mu)$ we obtain the renormalization
group relation
$$
g'= \frac{g+a_4(\mu';g,\mu)}{\left( 1+a_2(\mu';g,\mu) \right)^2}\;.
$$
The beta function is obtained by considering an infinitesimal
scale change and is given by
$$
\beta(g)=\mu'\partder{\mu'}
\left\{ a_4(\mu';g,\mu)-2g\,a_2(\mu';g,\mu) \right\}|_{\mu'=\mu}\,.
$$
Since the theory is renormalizable and infrared finite,
the two quantities $a_i(\mu';g,\mu)$ are functions of $g$ and the
ratio $\mu'/\mu$, thus and the beta function depends only on $g$.
At the one loop order we find $a^{(1)}_2$ independent of $\mu'$
and the beta function is obtained only from $a^{(1)}_4$ given in
\re{onel1}.
We find
$$
\beta^{(1)}(g)=
\frac {3}{32\pi^2}g^2 \mu'\partder{\mu'} \ln\frac{\mu'^2}{\mu^2}
=\frac {3}{16\pi^2}g^2 \,,
$$
which is the usual one loop result.

\eject

\eject
\newpage

\begin{figcap}

\item (a) Vertices $\G_{2n}(p_1,\cdots,p_{2n})$ and \\
(b) auxiliary vertices $\bG_{2n+2}(q,p_1,\cdots,p_{2n},q')$.

\item Graphical representation of the equation \re{auxv} defining
the auxiliary vertices $\bG_{2n+2}(q,p_1,\cdots,p_{2n},q')$.

\item Graphical representation of the auxiliary vertices
at zero loop.

\item Graphical representation of the leading contribution
of auxiliary vertices for $\L \to \infty$.

\item The leading contribution for $\L \to 0$ of auxiliary
vertices in which the pairs of momenta in the four point functions are
exceptional.

\item The leading contribution for $\L \to 0$ of derivatives
of auxiliary vertices in which all momenta are exceptional.

\end{figcap}

\pagestyle{empty}

\begin{figure}
\thicklines

\unitlength=1.50mm
\linethickness{0.4pt}
\begin{picture}(97.00,156.00)
\put(20.00,95.00){\framebox(15.00,6.00)[cc]{}}
\put(12.00,98.00){\vector(1,0){4.00}}
\put(16.00,98.00){\line(1,0){4.00}}
\put(43.00,98.00){\vector(-1,0){4.00}}
\put(39.00,98.00){\line(-1,0){4.00}}
\put(43.00,87.00){\vector(-1,1){4.00}}
\put(35.00,95.00){\line(1,-1){4.00}}
\put(12.00,87.00){\vector(1,1){4.00}}
\put(16.00,91.00){\line(1,1){4.00}}
\put(20.00,95.00){\line(0,0){0.00}}
\put(28.00,89.00){\makebox(0,0)[cc]{. . . . .}}
\put(12.00,83.00){\makebox(0,0)[cc]{$1$}}
\put(10.00,98.00){\makebox(0,0)[cc]{$q$}}
\put(44.00,98.00){\makebox(0,0)[lc]{$q'$}}
\put(43.00,83.00){\makebox(0,0)[lc]{$2n$}}
\put(54.00,98.00){\makebox(0,0)[cc]{$=$}}
\put(77.00,98.00){\oval(10.00,6.00)[]}
\put(64.00,98.00){\vector(1,0){4.00}}
\put(68.00,98.00){\line(1,0){4.00}}
\put(90.00,98.00){\vector(-1,0){4.00}}
\put(86.00,98.00){\line(-1,0){4.00}}
\put(66.00,87.00){\vector(1,1){4.00}}
\put(70.00,91.00){\line(1,1){4.00}}
\put(88.00,87.00){\vector(-1,1){4.00}}
\put(80.00,95.00){\line(1,-1){4.00}}
\put(95.00,98.00){\makebox(0,0)[cc]{$+$}}
\put(20.00,72.00){\makebox(0,0)[cc]{$\Sigma$}}
\put(39.00,72.00){\oval(10.00,6.00)[]}
\put(26.00,72.00){\vector(1,0){4.00}}
\put(30.00,72.00){\line(1,0){4.00}}
\put(28.00,61.00){\vector(1,1){4.00}}
\put(32.00,65.00){\line(1,1){4.00}}
\put(50.00,61.00){\vector(-1,1){4.00}}
\put(42.00,69.00){\line(1,-1){4.00}}
\put(64.00,69.00){\framebox(15.00,6.00)[cc]{}}
\put(87.00,72.00){\vector(-1,0){4.00}}
\put(83.00,72.00){\line(-1,0){4.00}}
\put(87.00,61.00){\vector(-1,1){4.00}}
\put(79.00,69.00){\line(1,-1){4.00}}
\put(56.00,61.00){\vector(1,1){4.00}}
\put(60.00,65.00){\line(1,1){4.00}}
\put(44.00,72.00){\vector(1,0){4.00}}
\put(48.00,72.00){\vector(1,0){12.00}}
\put(60.00,72.00){\line(1,0){4.00}}
\put(54.00,72.00){\circle*{2.00}}
\put(30.00,76.00){\makebox(0,0)[cc]{$q$}}
\put(54.00,76.00){\makebox(0,0)[cc]{$Q$}}
\put(83.00,76.00){\makebox(0,0)[lc]{$q'$}}
\put(87.00,57.00){\makebox(0,0)[lc]{$i_{2n}$}}
\put(50.00,57.00){\makebox(0,0)[lc]{$i_k$}}
\put(28.00,57.00){\makebox(0,0)[cc]{$i_1$}}
\put(28.00,46.00){\oval(10.00,6.00)[]}
\put(15.00,46.00){\vector(1,0){4.00}}
\put(19.00,46.00){\line(1,0){4.00}}
\put(17.00,35.00){\vector(1,1){4.00}}
\put(21.00,39.00){\line(1,1){4.00}}
\put(39.00,35.00){\vector(-1,1){4.00}}
\put(31.00,43.00){\line(1,-1){4.00}}
\put(2.00,46.00){\makebox(0,0)[cc]{$=$}}
\put(9.00,46.00){\makebox(0,0)[cc]{$\Sigma$}}
\put(54.00,46.00){\oval(10.00,6.00)[]}
\put(43.00,35.00){\vector(1,1){4.00}}
\put(47.00,39.00){\line(1,1){4.00}}
\put(65.00,35.00){\vector(-1,1){4.00}}
\put(57.00,43.00){\line(1,-1){4.00}}
\put(84.00,46.00){\oval(10.00,6.00)[]}
\put(73.00,35.00){\vector(1,1){4.00}}
\put(77.00,39.00){\line(1,1){4.00}}
\put(95.00,35.00){\vector(-1,1){4.00}}
\put(87.00,43.00){\line(1,-1){4.00}}
\put(33.00,46.00){\vector(1,0){4.00}}
\put(37.00,46.00){\line(1,0){12.00}}
\put(42.00,46.00){\circle*{2.00}}
\put(59.00,46.00){\vector(1,0){4.00}}
\put(69.00,46.00){\line(-1,0){6.00}}
\put(72.00,46.00){\makebox(0,0)[cc]{...}}
\put(79.00,46.00){\line(-1,0){5.00}}
\put(97.00,46.00){\vector(-1,0){4.00}}
\put(93.00,46.00){\line(-1,0){4.00}}
\put(93.00,50.00){\makebox(0,0)[lc]{$q'$}}
\put(84.00,31.00){\makebox(0,0)[cc]{$n_k$}}
\put(67.00,50.00){\makebox(0,0)[cc]{$Q_{n_1+n_2}$}}
\put(67.00,46.00){\circle*{2.00}}
\put(54.00,31.00){\makebox(0,0)[cc]{$n_2$}}
\put(42.00,51.00){\makebox(0,0)[cc]{$Q_{n_1}$}}
\put(28.00,31.00){\makebox(0,0)[cc]{$n_1$}}
\put(19.00,51.00){\makebox(0,0)[cc]{$q$}}
\put(54.00,21.00){\makebox(0,0)[lc]{Fig. 2}}
\put(13.00,72.00){\makebox(0,0)[rc]{$+$}}
\put(95.00,72.00){\makebox(0,0)[cc]{$=$}}
\put(39.00,63.00){\makebox(0,0)[cc]{. . . . .}}
\put(72.00,63.00){\makebox(0,0)[cc]{. . . . .}}
\put(84.00,37.00){\makebox(0,0)[cc]{. . . . .}}
\put(54.00,37.00){\makebox(0,0)[cc]{. . . . .}}
\put(28.00,37.00){\makebox(0,0)[cc]{. . . . .}}
\put(69.00,138.00){\framebox(15.00,6.00)[cc]{}}
\put(61.00,141.00){\vector(1,0){4.00}}
\put(65.00,141.00){\line(1,0){4.00}}
\put(92.00,141.00){\vector(-1,0){4.00}}
\put(88.00,141.00){\line(-1,0){4.00}}
\put(92.00,130.00){\vector(-1,1){4.00}}
\put(84.00,138.00){\line(1,-1){4.00}}
\put(61.00,130.00){\vector(1,1){4.00}}
\put(65.00,134.00){\line(1,1){4.00}}
\put(69.00,138.00){\line(0,0){0.00}}
\put(15.00,142.00){\circle{10.00}}
\put(26.00,154.00){\vector(-1,-1){4.00}}
\put(22.00,150.00){\line(-1,-1){4.00}}
\put(4.00,154.00){\vector(1,-1){4.00}}
\put(8.00,150.00){\line(1,-1){4.00}}
\put(4.00,130.00){\vector(1,1){4.00}}
\put(8.00,134.00){\line(1,1){4.00}}
\put(26.00,130.00){\vector(-1,1){4.00}}
\put(22.00,134.00){\line(-1,1){4.00}}
\put(59.00,128.00){\makebox(0,0)[cc]{$1$}}
\put(59.00,141.00){\makebox(0,0)[cc]{$q$}}
\put(93.00,141.00){\makebox(0,0)[lc]{$q'$}}
\put(2.00,128.00){\makebox(0,0)[cc]{$2$}}
\put(2.00,156.00){\makebox(0,0)[cc]{$1$}}
\put(28.00,156.00){\makebox(0,0)[lc]{$2n$}}
\put(94.00,128.00){\makebox(0,0)[lc]{$2n$}}
\put(15.00,132.00){\makebox(0,0)[cc]{ . . . . .}}
\put(77.00,132.00){\makebox(0,0)[cc]{. . . . .}}
\put(77.00,125.00){\makebox(0,0)[cc]{Fig. 1b}}
\put(15.00,125.00){\makebox(0,0)[cc]{Fig. 1a}}
\put(11.00,137.00){\line(1,1){1.00}}
\put(19.00,137.00){\line(-1,1){1.00}}
\put(20.00,148.00){\line(-1,-1){2.00}}
\put(10.00,148.00){\line(1,-1){2.00}}
\end{picture}

\end{figure}

\begin{figure}
\thicklines

\unitlength=1.50mm
\linethickness{0.4pt}
\begin{picture}(108.00,157.00)
\put(12.00,57.00){\makebox(0,0)[cc]{$q$}}
\put(10.00,55.00){\line(1,0){6.00}}
\put(18.00,55.00){\circle{4.00}}
\put(28.00,55.00){\line(-1,0){8.00}}
\put(30.00,55.00){\circle{4.00}}
\put(40.00,55.00){\line(-1,0){8.00}}
\put(42.00,55.00){\circle{4.00}}
\put(44.00,55.00){\line(1,0){8.00}}
\put(54.00,55.00){\circle{4.00}}
\put(56.00,55.00){\line(1,0){8.00}}
\put(66.00,55.00){\circle{4.00}}
\put(68.00,55.00){\line(1,0){6.00}}
\put(72.00,57.00){\makebox(0,0)[lc]{$-q$}}
\put(67.00,53.00){\line(1,-3){2.00}}
\put(63.00,47.00){\line(1,3){2.00}}
\put(55.00,53.00){\line(1,-3){2.00}}
\put(51.00,47.00){\line(1,3){2.00}}
\put(43.00,53.00){\line(1,-3){2.00}}
\put(39.00,47.00){\line(1,3){2.00}}
\put(31.00,53.00){\line(1,-3){2.00}}
\put(29.00,53.00){\line(-1,-3){2.00}}
\put(19.00,53.00){\line(1,-3){2.00}}
\put(17.00,53.00){\line(-1,-3){2.00}}
\put(27.00,45.00){\makebox(0,0)[cc]{$p$}}
\put(33.00,45.00){\makebox(0,0)[cc]{$p_i$}}
\put(51.00,45.00){\makebox(0,0)[cc]{$-p$}}
\put(57.00,45.00){\makebox(0,0)[cc]{$p_j$}}
\put(50.00,37.00){\makebox(0,0)[cc]{Fig. 6}}
\put(3.00,120.00){\makebox(0,0)[cc]{$\Sigma$}}
\put(13.00,120.00){\vector(1,0){4.00}}
\put(17.00,120.00){\line(1,0){32.00}}
\put(14.00,122.00){\makebox(0,0)[cc]{$q$}}
\put(19.00,112.00){\vector(1,2){2.00}}
\put(21.00,116.00){\line(1,2){2.00}}
\put(27.00,112.00){\vector(-1,2){2.00}}
\put(25.00,116.00){\line(-1,2){2.00}}
\put(31.00,112.00){\vector(1,2){2.00}}
\put(33.00,116.00){\line(1,2){2.00}}
\put(39.00,112.00){\vector(-1,2){2.00}}
\put(37.00,116.00){\line(-1,2){2.00}}
\put(52.00,112.00){\vector(1,2){2.00}}
\put(54.00,116.00){\line(1,2){2.00}}
\put(60.00,112.00){\vector(-1,2){2.00}}
\put(58.00,116.00){\line(-1,2){2.00}}
\put(45.00,120.00){\line(1,0){17.00}}
\put(66.00,120.00){\vector(-1,0){4.00}}
\put(65.00,122.00){\makebox(0,0)[cc]{$-q$}}
\put(46.00,114.00){\makebox(0,0)[cc]{. . . . .}}
\put(60.00,109.00){\makebox(0,0)[cc]{$i_{2n}$}}
\put(19.00,109.00){\makebox(0,0)[cc]{$i_1$}}
\put(27.00,109.00){\makebox(0,0)[cc]{$i_2$}}
\put(23.00,120.00){\circle*{2.00}}
\put(35.00,120.00){\circle*{2.00}}
\put(56.00,120.00){\circle*{2.00}}
\put(104.00,120.00){\makebox(0,0)[lc]{$=$  $\gamma_4(\Lambda)$}}
\put(89.00,124.00){\line(1,-1){8.00}}
\put(89.00,116.00){\line(1,1){8.00}}
\put(93.00,120.00){\circle*{2.00}}
\put(50.00,101.00){\makebox(0,0)[cc]{Fig. 4}}
\put(16.00,85.00){\circle{4.00}}
\put(14.00,85.00){\line(-1,0){8.00}}
\put(18.00,85.00){\line(1,0){10.00}}
\put(30.00,85.00){\circle{4.00}}
\put(32.00,85.00){\line(1,0){10.00}}
\put(44.00,85.00){\circle{4.00}}
\put(52.00,85.00){\line(-1,0){6.00}}
\put(57.00,85.00){\makebox(0,0)[cc]{. . . . . }}
\put(62.00,85.00){\line(1,0){6.00}}
\put(73.00,85.00){\oval(10.00,4.00)[]}
\put(78.00,85.00){\line(1,0){8.00}}
\put(76.00,83.00){\line(1,-3){2.00}}
\put(74.00,83.00){\line(1,-6){1.00}}
\put(70.00,83.00){\line(-1,-3){2.00}}
\put(72.00,83.00){\line(-1,-6){1.00}}
\put(50.00,70.00){\makebox(0,0)[cc]{Fig. 5}}
\put(15.00,83.00){\line(-1,-3){2.00}}
\put(17.00,83.00){\line(1,-3){2.00}}
\put(29.00,83.00){\line(-1,-3){2.00}}
\put(31.00,83.00){\line(1,-3){2.00}}
\put(43.00,83.00){\line(-1,-3){2.00}}
\put(45.00,83.00){\line(1,-3){2.00}}
\put(3.00,153.00){\vector(1,0){4.00}}
\put(7.00,153.00){\line(1,0){4.00}}
\put(11.00,149.00){\framebox(15.00,8.00)[cc]{$0$}}
\put(34.00,153.00){\vector(-1,0){4.00}}
\put(30.00,153.00){\line(-1,0){4.00}}
\put(26.00,153.00){\line(0,0){0.00}}
\put(34.00,141.00){\vector(-1,1){4.00}}
\put(30.00,145.00){\line(-1,1){4.00}}
\put(3.00,141.00){\vector(1,1){4.00}}
\put(7.00,145.00){\line(1,1){4.00}}
\put(4.00,155.00){\makebox(0,0)[cc]{$q$}}
\put(2.00,138.00){\makebox(0,0)[cc]{$1$}}
\put(19.00,143.00){\makebox(0,0)[cc]{. . . . .}}
\put(35.00,138.00){\makebox(0,0)[cc]{$2n$}}
\put(33.00,155.00){\makebox(0,0)[cc]{$q'$}}
\put(41.00,153.00){\makebox(0,0)[cc]{$=$}}
\put(48.00,153.00){\makebox(0,0)[cc]{$\Sigma$}}
\put(55.00,153.00){\vector(1,0){4.00}}
\put(59.00,153.00){\line(1,0){32.00}}
\put(57.00,155.00){\makebox(0,0)[cc]{$q$}}
\put(61.00,145.00){\vector(1,2){2.00}}
\put(63.00,149.00){\line(1,2){2.00}}
\put(69.00,145.00){\vector(-1,2){2.00}}
\put(67.00,149.00){\line(-1,2){2.00}}
\put(73.00,145.00){\vector(1,2){2.00}}
\put(75.00,149.00){\line(1,2){2.00}}
\put(81.00,145.00){\vector(-1,2){2.00}}
\put(79.00,149.00){\line(-1,2){2.00}}
\put(94.00,145.00){\vector(1,2){2.00}}
\put(96.00,149.00){\line(1,2){2.00}}
\put(102.00,145.00){\vector(-1,2){2.00}}
\put(100.00,149.00){\line(-1,2){2.00}}
\put(87.00,153.00){\line(1,0){17.00}}
\put(108.00,153.00){\vector(-1,0){4.00}}
\put(107.00,155.00){\makebox(0,0)[cc]{$q'$}}
\put(88.00,147.00){\makebox(0,0)[cc]{. . . . .}}
\put(102.00,142.00){\makebox(0,0)[cc]{$i_{2n}$}}
\put(61.00,142.00){\makebox(0,0)[cc]{$i_1$}}
\put(69.00,142.00){\makebox(0,0)[cc]{$i_2$}}
\put(50.00,132.00){\makebox(0,0)[cc]{Fig. 3}}
\put(8.00,87.00){\makebox(0,0)[cc]{$q$}}
\put(84.00,87.00){\makebox(0,0)[cc]{$-q$}}
\end{picture}

\end{figure}

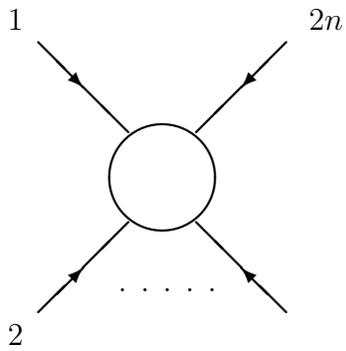
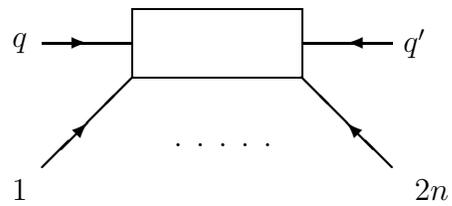
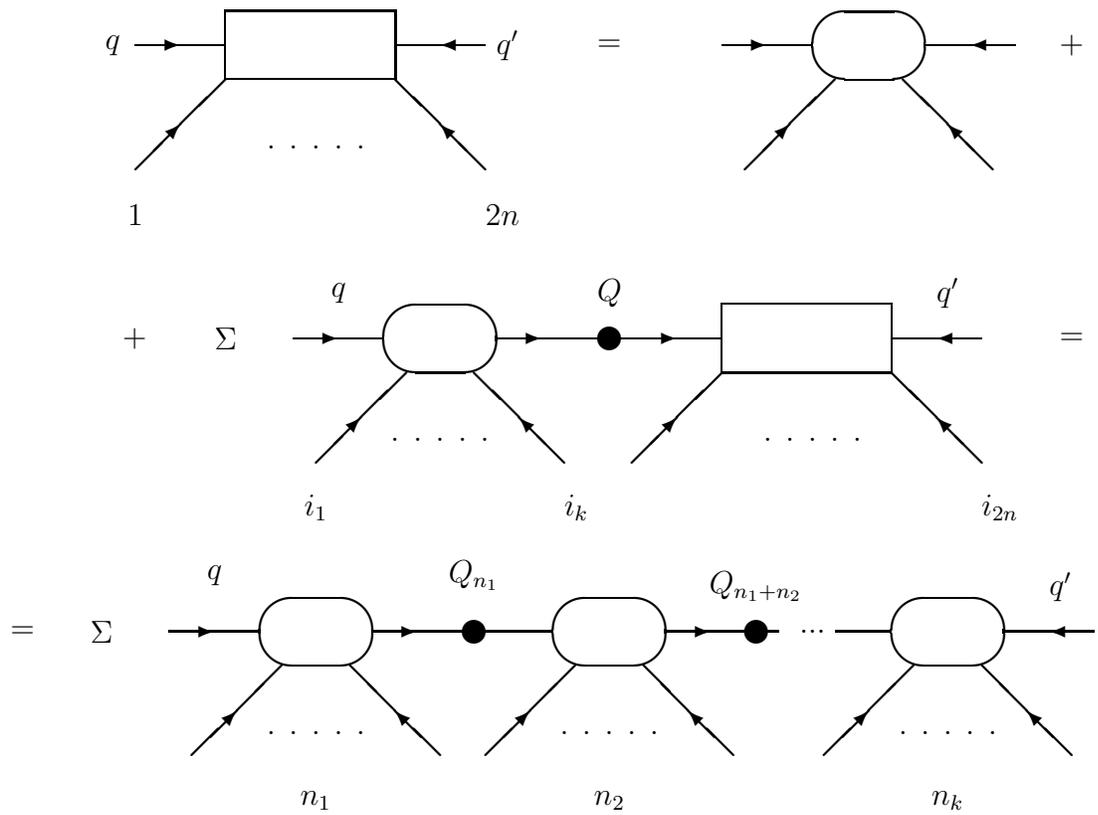
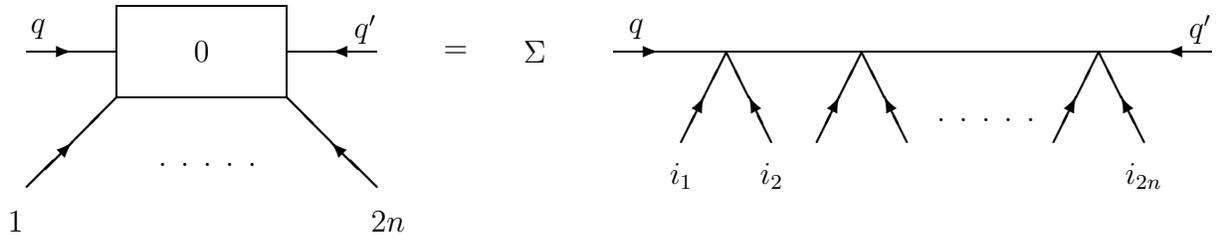
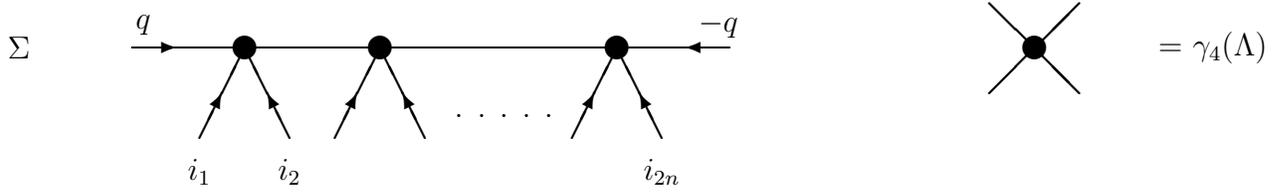
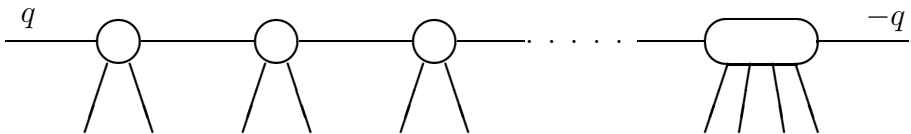
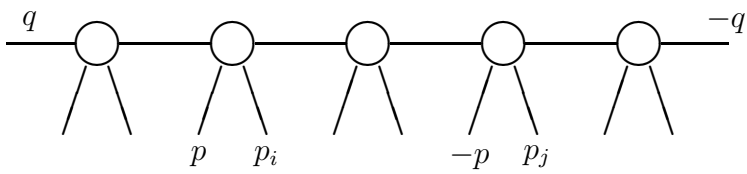
\end{document}